\documentclass[12pt]{article}

\begin{document}

\thispagestyle{empty}
\renewcommand{\thefootnote}{\fnsymbol{footnote}}

\begin{flushright}
{\small
SLAC--PUB--7752\\
February 1998\\}
\end{flushright}

\vspace{.8cm}

\begin{center}
{\bf\large
LATTICE AND COMPENSATION SCHEMES \\
FOR THE PEP-II INTERACTION REGION
\footnote{Work supported by Department of 
Energy contract  DE--AC03--76SF00515.}}

\vspace{1cm}

Y. Nosochkov, Y. Cai, M.H.R. Donald, J. Irwin \\
D.M. Ritson, J. Seeman, M. Sullivan \\ 
Stanford Linear Accelerator Center, Stanford University,
Stanford, CA  94309\\

%\medskip
\end{center}

\vfill

\begin{center}
{\bf\large
Abstract }
\end{center}

\begin{quote}
The PEP-II interaction region is designed to accommodate asymmetric beam 
energies, head-on collisions, small bunch spacing and to provide
low $\beta^{*}$ for high luminosity. Local correction schemes are 
implemented to compensate non-linear chromaticity from the IP doublets 
as well as coupling, orbit and focusing effects from the 6 Tm asymmetric 
detector solenoid. The main IR optics features and local 
correction schemes are presented. MAD\cite{mad} code is used for 
the optics calculations.
\end{quote}

\vfill

\begin{center}
{\it Talk presented at}
{\it Advanced ICFA Workshop on Beam Dynamics Issues for e+e-
Factories}\\
{\it LNF--INFN, Frascati, Italy}\\
{\it October 20--October 25, 1997}\\
\end{center}

\newpage

\pagestyle{plain}

\section{IR optics}
In the PEP-II asymmetric collider the High Energy Ring (HER)
and the Low Energy Ring (LER) consist of 6 arcs and 6 straight 
sections\cite{her,ler}. The two rings are vertically separated
except at the interaction point (IP) where the 9 GeV electron
and 3.1 GeV positron beams are brought into collision.

The PEP-II interaction region (IR) optics has to meet the requirement
for high luminosity, provide adequate dynamic aperture and satisfy
geometric constraints. The following IR conditions are applied
to attain the design luminosity: low IP beta 
($\beta^{*}_{y}=0.015$ m), zero IP dispersion, and head-on collisions.
Head-on collisions minimize the effect of synchro-betatron
resonances, though they require a more complicated separation scheme 
compared to a crossing angle collision. Maximizing dynamic aperture
requires: compensation of the 6 Tm detector solenoid, compensation of
non-linear chromaticity produced by the IP doublets, matching IR
optics to the arcs, and maximizing beam separation at parasitic 
crossing points. Geometric constraints include: separation of two 
ring components after collision, providing 0.89 m vertical
separation between the LER and HER after the IP, matching the IR
trajectory to the arcs, and fitting the IR components into the
existing tunnel.

\begin{figure}[h]
  \includegraphics{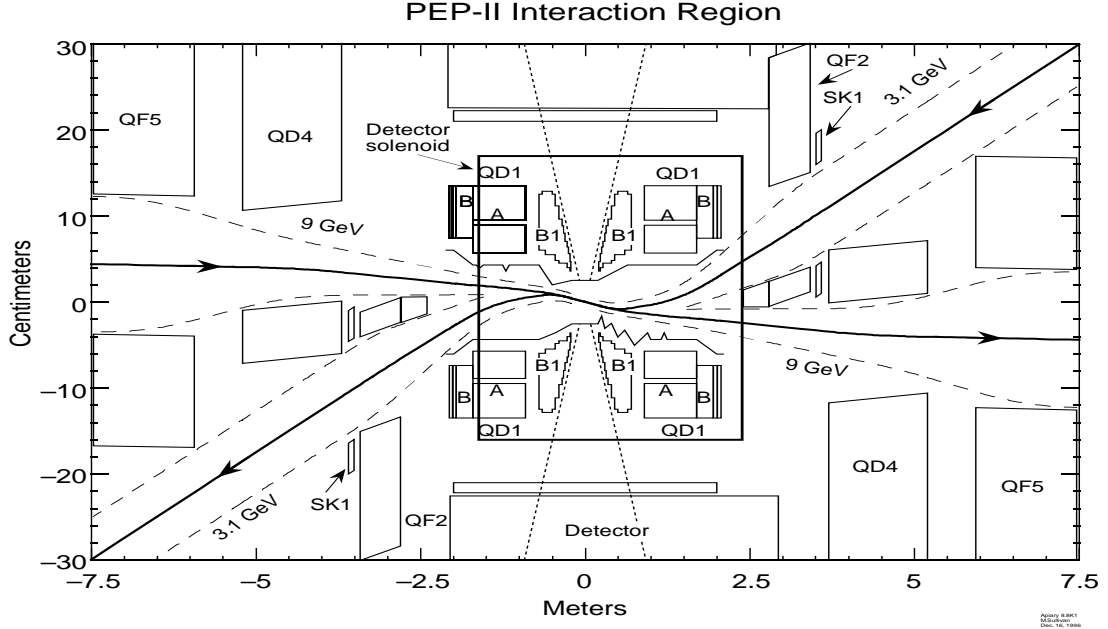}
  \vspace{9cm}
  \caption{\it
    Top view of the PEP-II magnets and detector solenoid near IP.
  \label{ip} }
\end{figure}

The PEP-II IR optics has the following symmetry with respect to 
the IP: symmetric longitudinal positions of the ring components,
symmetric quadrupole focusing, symmetric vertical bending, 
and antisymmetric horizontal bending. The IR focusing 
symmetry maintains the ring symmetry with respect to the IP 
and allows the IR sextupole correction to be symmetric.
The nominal $\beta^{*}_{x}/\beta^{*}_{y}$ values at the IP for 
the design luminosity are 0.5 m/1.5 cm for the LER and 0.667 m/2 cm 
for the HER. Because of the asymmetric beam energies the IR optics
is essentially independent for the two rings which requires a fast 
separation of the two ring components after the IP. The first magnets
near the collision point are horizontal separation bends B1 placed 
at $s=\pm 0.21$ m from the IP. The asymmetry in beam energies is just 
enough to produce sufficient beam separation in the B1 for a zero 
crossing angle. To minimize the size of the divergent beams after the 
IP the LER $\beta$ functions are focused with the quadrupole doublet 
QD1, QF2 placed next to the B1 (see Fig.\ref{ip}).
The B1 and QD1 are the only magnets shared by the two rings.
The QD1 produces helpful vertical focusing on the HER beam as well,
and the focusing is completed with a separate HER doublet QD4, QF5.
Besides the field gradient the QD1, QF2, QD4 magnets have a 
vertical dipole field on the reference orbit of one or the other
beam to help the horizontal beam separation. As a result, the two 
beams are sufficiently separated at the first two parasitic collisions: 
$\Delta x=3.3$ mm $(11\sigma_{x})$ at $s=\pm 0.63$ m and
$\Delta x=18$ mm $(35\sigma_{x})$ at $s=\pm 1.26$ m. The QF2, QD4, QF5
are septum magnets to make easier separation of the LER and HER
components.

The antisymmetric horizontal bending in the IR allows us to minimize 
the IR trajectory excursions by adjustment of the horizontal orbit 
slope at the IP with respect to the tunnel. The optimized value 
of this angle is -16.9 mrad. The vertical separation between the 
rings is achieved with a pair of LER vertical bends BV1,2 on either side 
of the IP. The bends are placed $2\pi$ apart in $y$-phase and have 
identical $\beta_{y}$ to cancel vertical dispersion. One complication 
of the IR geometry in the LER is the interleaved horizontal and vertical 
bends which cause a tilt of the beam eigenplanes with respect to the 
mid-plane. Once the IR magnets are properly aligned to close the 
ring circumference, the above leads to a small betatron coupling 
which has to be canceled.

\begin{figure}[h]
  \includegraphics{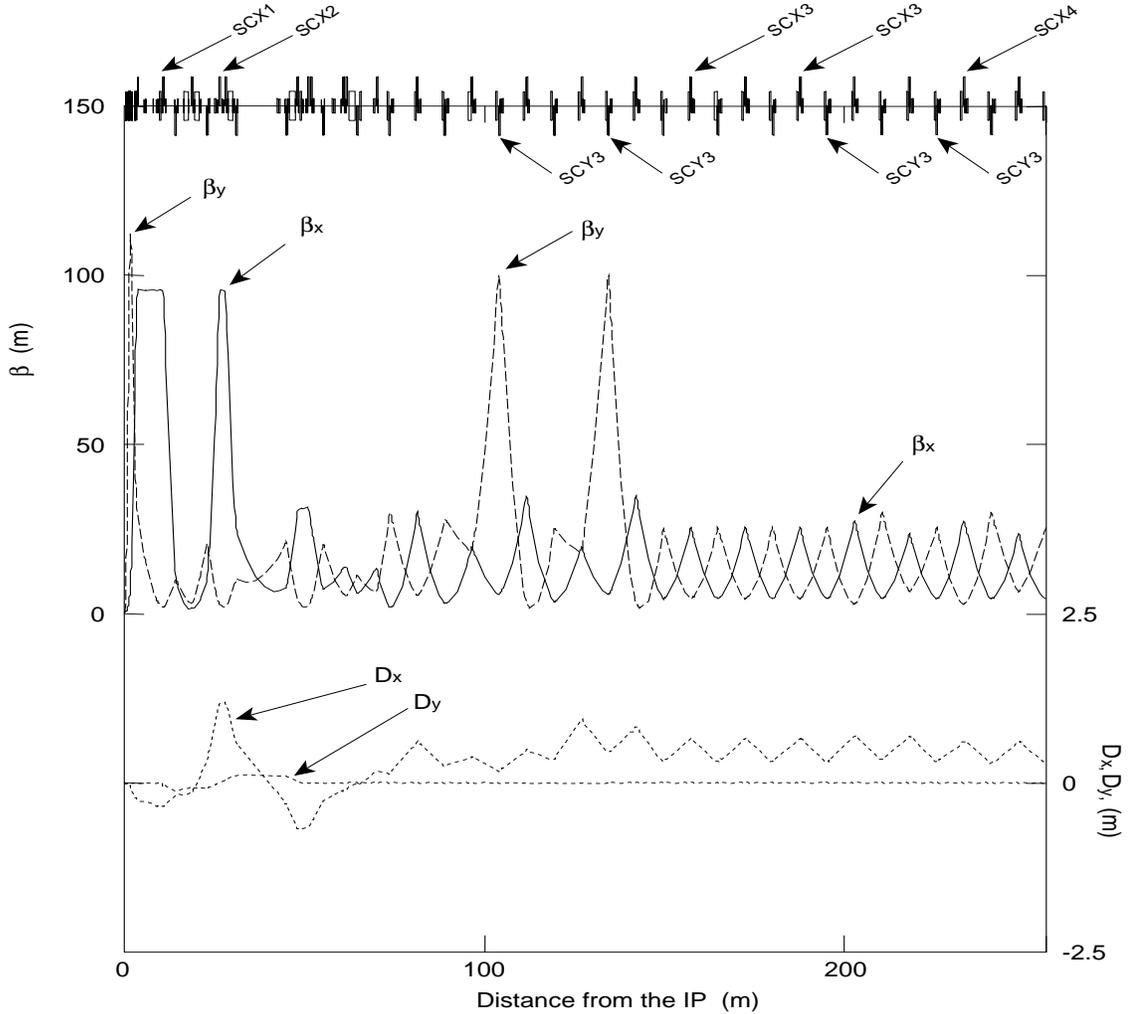}
  \vspace{14cm}
  \caption{\it
    Half IR in the LER with local and semi-local sextupoles.
  \label{hir} }
\end{figure}

Besides the vertical bends the IR in the LER accommodates a pair of 
local sextupoles to correct $x$-chromaticity generated by the 
IP doublet. This requires a $-I$ transformation between the sextupoles, 
a minimum $x$-phase advance between the sextupoles and the doublet, 
a high $\beta_{x}/\beta_{y}$ ratio and a non-zero dispersion at the 
sextupoles. A similar requirement for a high $\beta_{x}/\beta_{y}$ 
or $\beta_{y}/\beta_{x}$ ratio is applied to the semi-local sextupoles 
placed in the nearby arcs where local $\beta$ bumps were created. One 
half of the IR and the adjacent arc in the LER is shown in 
Fig.\ref{hir} where the IP is at $s=0$. The HER has similar $\beta$ bumps 
in the arcs near the IR for the semi-local sextupoles, but the IR optics
is much simpler compared to the LER.
\section{Local and semi-local chromatic correction}
In a low $\beta$ optics the $\beta$ functions at the IP doublets
attain very high values and the doublet quads are stronger 
than other ring quadrupoles. This results in two single sources 
of large chromaticity. In the PEP-II lattice the two IP doublets 
make about $20\%$ to $30\%$ of the ring linear chromaticity. 
For such a large chromatic perturbation
the non-linear part becomes significant. Moreover, the 
non-linear chromaticity from the two doublets amplify each other, 
while in the rest of the ring it tends to cancel out because of 
periodicity or special phase advance such as $90^{o}$ per cell. 
The chromatic effect of the IP doublets can significantly reduce 
the momentum dependent dynamic aperture and requires special 
correction. We refer to a local compensation when correcting 
sextupoles are placed close to the IP doublet, and to a 
semi-local correction if they are located farther from the 
doublet in betatron phase.

\begin{figure}[t]
  \includegraphics{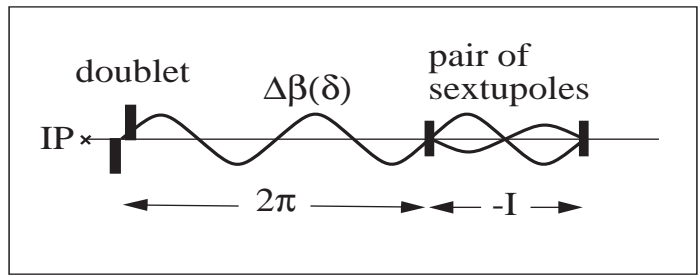}
  \vspace{4.2cm}
  \caption{\it
   Chromatic doublet correction with a pair of sextupoles.
  \label{chrom} }
\end{figure}

The method to use the correcting sextupoles is: 
1) to place them in a nearby dispersive region with the 
phase advance in the corresponding plane of
$\pi \times integer + \Delta \mu$ from the doublet, where 
$\Delta \mu$ is a fine adjustment for optimum correction; 
2) to use pairs of identical sextupoles separated by a $-I$ 
transformation to cancel sextupole geometric aberrations;
3) to have large $\beta_{x}/\beta_{y}$ or $\beta_{y}/\beta_{x}$
ratio at the sextupoles for orthogonal $x$ and $y$ correction 
and to minimize sextupole strengths; 
4) to have no other sextupoles within each sextupole pair
to minimize octupole-like tune shift with amplitude.
It is advantageous to place sextupoles as close as possible to 
the corrected doublet to avoid a disturbing effect by the magnets
in the middle. The chromatic effects to be minimized are tune 
shift with momentum; variation of $\beta$ function with 
momentum at IP, injection point and RF cavities; higher 
order dispersion at IP and RF cavities. It is also necessary
to keep to a minimum the sextupole geometric aberrations 
such as tune shift with amplitude and resonance driving terms. 
For a better correction all ring sextupole families have to be
optimized. Fig.\ref{chrom} shows how chromatic beta perturbation 
$\Delta \beta (\delta)$ produced by a doublet is compensated by 
the opposite beta wave from a pair of sextupoles.

In the HER the $60^{o}$ cells allow us to place 2
($x$ and $y$) semi-local non-interleaved sextupole pairs in 
each arc adjacent to the IR. A local $\beta$ bump was created 
in these arcs to increase $\beta$ ratio at the sextupoles 
from 3 to 13. The rest of the HER is corrected with standard 
two family interleaved sextupoles SF, SD placed in the other 4 arcs.
In the LER the correction scheme per half IR includes one local
pair of $x$-sextupoles placed next to the IP doublet in a 
$-I$ section with non-zero dispersion and high $\beta$ ratio, and
4 semi-local non-interleaved sextupole pairs in the arc near IR
(see Fig.\ref{hir}).
The $90^{o}$ cells in the LER allow more room for the
sextupole pairs compared to the HER. A $\beta$ bump 
in the two arcs increases the $\beta$ ratio at the sextupoles from 
5.8 to 14. The rest of the LER has 4 non-interleaved pairs of
SF1, SF2 or SD1, SD2 sextupoles per arc to correct linear chromaticity.
In both rings the correction scheme is symmetric about IP.
The IR local chromatic compensation significantly reduces momentum
dependent tune shift and variation of $\beta (\delta)$.
\section{Solenoid compensation}
\begin{figure}[h]
  \includegraphics{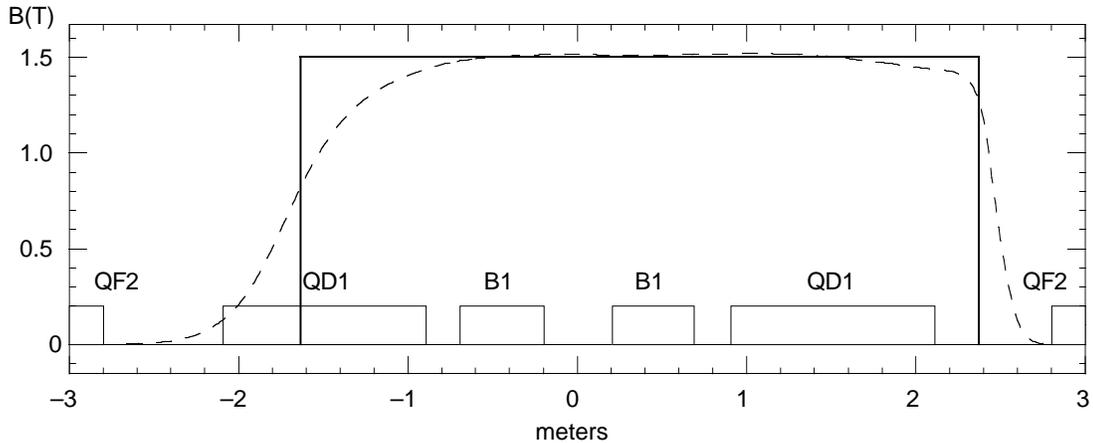}
  \vspace{6cm}
  \caption{\it
    Hard (solid) and soft (dash) model of the solenoid field profile.
  \label{profile} }
\end{figure}

The PEP-II detector solenoid has a significant effect on the beam
optics. The integrated solenoid field is 6.07 Tm which
rotates the LER beam eigenplanes by $17^{o}$. The most part of the 
solenoid field is located within 4 m near IP with the maximum 
field of 1.5 T. With the fringe field included the solenoid length 
extends over 6 m. The solenoid field profile is shown in 
Fig.\ref{profile}. Other complications are:
1) the solenoid center is placed 37 cm from the IP in the direction 
of the HER beam;
2) the solenoid overlaps the B1, QD1 and QF2 magnets on both sides 
of the IP, hence the fields are superimposed;
3) neither beam is parallel to the solenoid axis, and the solenoid 
is horizontally tilted with respect to the beam orbit at the IP
(see Fig.\ref{ip}).
The solenoid effects are: 
1) coupling of $x$ and $y$ betatron motion;
2) beam focusing in both planes;
3) vertical orbit and dispersion caused by the solenoid tilt,
and horizontal orbit and dispersion induced by the coupling.

\begin{figure}[b]
  \includegraphics{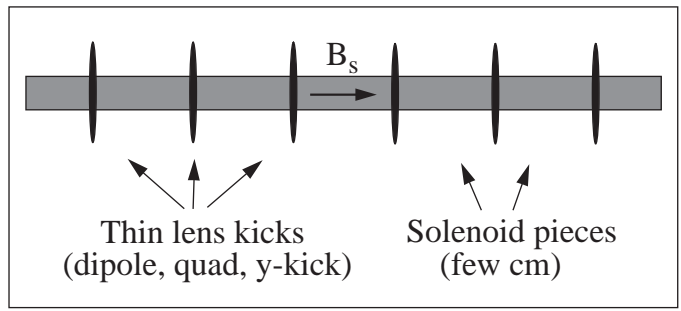}
  \vspace{4.7cm}
  \caption{\it
    Field superposition model.
  \label{model} }
\end{figure}

To model the superposition of the solenoid, quadrupole and dipole fields 
in the optics and tracking codes each of the B1, QD1 and QF2 magnets 
have been replaced by a combination of thin lenses with thick solenoid 
pieces between them. In the $hard$ edge model the solenoid field is 
held constant, the solenoid pieces are aligned along the beam and the 
vertical orbit from the solenoid tilt is simulated with a set of thin 
vertical kicks. In the $soft$ edge model the field is a piece-wise
function of the longitudinal position, and each solenoid piece is 
properly aligned with respect to the beam trajectory.
The field model is illustrated in Fig.\ref{model}.

The solenoid correction requirements include:
1) uncoupled $x$ and $y$ betatron motion at the IP;
2) no residual orbit or dispersion at the IP;
3) nominal $\beta$ function at the IP;
4) no optics perturbation outside the IR.
With the asymmetric solenoid this implies local and independent correction
on the left and right sides of the IP. In particular, the transfer matrix
from the arcs to the IP must be uncoupled and matched independently 
on each side of the IR.

The PEP-II design makes it difficult to use compensating solenoids near 
the IP to correct coupling from the detector solenoid, therefore we have 
adopted a skew quadrupole correction system\cite{solen}. The correction 
scheme per each half IR consists of:
1) 4 skew quadrupoles to uncouple the transfer matrix between the arcs and
the IP;
2) 2 skew quadrupoles placed in dispersive regions to correct vertical
dispersion $\eta_{y}$ and slope $\eta'_{y}$ at the IP;
3) 2 horizontal and 2 vertical orbit correctors to correct the orbit at 
the IP;
4) 8 quadrupoles or more to match twiss functions and horizontal
dispersion $\eta_{x}$, $\eta'_{x}$ at the IP.
The optimum phases for the 4 skew quadrupoles to correct coupling are
$[\mu_{x},\mu_{y}]=[0,0],[0,\pi /2],[\pi /2,0],[\pi /2,\pi /2]$ 
($mod$ $\pi$) from the IP. The optimum phases for orbit and dispersion 
correctors are 0 and $\pi /2$ ($mod$ $\pi$) from the IP in the corresponding
plane. The following adjustments helped to minimize the orbit excursions
near the IP:
1) solenoid horizontal tilt angle of 20.4 mrad with respect to the beam
trajectory at the IP;
2) vertical displacement of the IP by 3.7 mm;
3) vertical displacement of QF2, QD4 quadrupoles by a few mm to help
orbit correction.

In the existing lattice it is not always possible to find the absolutely
optimum positions for all correctors, therefore the coupling, orbit, 
dispersion and $\beta$ function corrections are not completely 
independent and for exact correction require simultaneous adjustment 
of all the above correctors. However, for a small variation of individual 
optics parameters a smaller number of correctors can be used leaving 
practically negligible residual perturbation.
\end{document}